\documentclass[11pt]{article}
\usepackage{amsmath,amsfonts}

\begin{document}

\title{\textbf{Renormalizability of a generalized gauge fixing interpolating %
among the Coulomb, Landau and maximal Abelian gauges} }
\author{\textbf{M.A.L. Capri$^{a}$\thanks{%
marcio@dft.if.uerj.br}} \ , \textbf{R.F. Sobreiro}$^{a}$\thanks{%
sobreiro@uerj.br} \ , \textbf{S.P. Sorella}$^{a}$\thanks{%
sorella@uerj.br}{\ }\footnote{Work supported by FAPERJ, Funda{\c
c}{\~a}o de Amparo {\`a} Pesquisa do Estado do Rio de Janeiro,
under the program {\it Cientista do Nosso Estado},
E-26/151.947/2004.} \ , \textbf{R. Thibes}$^{a}$\thanks{%
thibes@dft.if.uerj.br} \\\\
\textit{$^{a}$\small{Departamento de F\'{\i }sica Te\'{o}rica}}\\
\textit{\small{Instituto de F\'{\i }sica, UERJ, Universidade do Estado do Rio de Janeiro}} \\
\textit{\small{Rua S{\~a}o Francisco Xavier 524, 20550-013 Maracan{\~a}}} \\
\textit{\small{Rio de Janeiro, Brasil}}}
\date{}
\maketitle

\begin{abstract}

\noindent A detailed discussion of the renormalization properties
of a class of gauges which interpolates among the Landau, Coulomb
and Maximal Abelian gauges is provided in the framework of the
algebraic renormalization in Euclidean Yang-Mills theories in four
dimensions.

\end{abstract}

\newpage

\section{Introduction}

Interpolating gauges in Yang-Mills theories have been used for
renormalizability purposes and in order to understand the behavior
of gauge invariant operators
\cite{Baulieu:1998kx,Dudal:2004rx,Fischer:2005qe,Capri:2005zj}. In
\cite{Baulieu:1998kx}, a gauge fixing interpolating between the
Coulomb gauge and the Landau gauge has been discussed and its
renormalizability established. Further, a gauge which interpolates
between the Landau and the maximal Abelian gauge (MAG) was
constructed in \cite{Dudal:2004rx}. In this work this
interpolating gauge was used in order to study the vacuum energy
in the MAG. It has been shown in
\cite{Baulieu:1998kx,Dudal:2004rx} that those two types of
interpolating gauges are renormalizable to all orders in
perturbation theory. A generalization which connects these two
class of gauges was proposed and analyzed at the classical level
in \cite{Capri:2005zj}, providing thus a gauge which interpolates
between the Coulomb, the Landau and the maximal Abelian
gauges (the CLM gauge). \\\\
We point out that these three gauges (Coulomb, Landau, MAG) have
been used to understand specific aspects of the nonperturbative
infrared region of Yang-Mills theories, from theoretical as well
as from lattice numerical simulations and phenomenological point
of views. Therefore, a generalized gauge fixing interpolating
among all these three gauges might be helpful in order to achieve
a unifying picture of the
behavior of gauge invariant quantities like, for instance, the vacuum energy.\\\\
Remarkably, the three gauges discussed here can be obtained
through the minimization of a suitable functional, a feature which
allows to construct a lattice formulation of these gauge
conditions. Not surprisingly, the gauge fixing which interpolates
among those three gauges turns out, in a suitable limit, to be
defined as a minimization of an interpolating functional, making
possible the implementation of a lattice formulation, see Appendix
\ref{apA}.\\\\In this work we discuss the quantum aspects of the
interpolating gauge, our main result being that of establishing
the all orders multiplicative renormalizability of this
generalized gauge. In section 2 we review the construction of the
CLM interpolating gauge introduced in \cite{Capri:2005zj}. In
section 3 we explain how to control the breaking of the Lorentz
invariance due to the gauge fixing. The proof of the
multiplicative renormalizability is given in section 4. Finally,
in section 5 we provide our conclusions.

\section{Interpolating gauge fixing}

In order to introduce the Coulomb-Landau-Maximal Abelian (CLM)
interpolating gauge let us briefly fix the notation. According to
the notation used in \cite{Dudal:2004rx}, we decompose the gauge
field $A_{\mu }^{A}$, $A\;\in\;\{1,\ldots,N^{2}-1\}$, into
off-diagonal and diagonal components, namely
\begin{equation}
A_{\mu }^{A}T^{A}=A_{\mu }^{a}T^{a}+A_{\mu }^{i}T^{i}\;,\label{mag1}
\end{equation}
where $T^{A}$ are the anti-hermitian, $T^\dagger=-T$, generators
of the gauge group $SU(N)$, $\left[T^{A},T^{B}\right]
=f^{ABC}T^{C}$. The indices $\{i,j\}$ label the $N-1$ diagonal
generators of the Cartan subalgebra. The remaining $N(N-1)$
off-diagonal generators will be labelled by the indices
$a,b,c,\ldots $.\\\\ For the nilpotent $BRST$ transformations of
the fields, we have
\begin{eqnarray}
s_gA_{\mu }^{a}\!\!\!\! &=&\!\!\!\!-(D_{\mu
}^{ab}c^{b}+gf^{abc}A_{\mu }^{b}c^{c}+gf^{abi}A_{\mu
}^{b}c^{i})\;,  \nonumber \\
s_gA_{\mu }^{i}\!\!\!\! &=&\!\!\!\!-(\partial _{\mu
}c^{i}+gf^{abi}A_{\mu
}^{a}c^{b})\;,  \nonumber \\
s_gc^{a}\!\!\!\!
&=&\!\!\!\!gf^{abi}c^{b}c^{i}+\frac{g}{2}f^{abc}c^{b}c^{c}\;,
\nonumber \\
s_gc^{i}\!\!\!\! &=&\!\!\!\!\frac{g}{2}f^{abi}c^{a}c^{b}\;,  \nonumber \\
s_g\bar{c}^{a}\!\!\!\! &=&\!\!\!\!b^{a},\qquad s_gb^{a}=0\;,  \nonumber \\
s_g\bar{c}^{i}\!\!\!\! &=&\!\!\!\!b^{i},\qquad s_gb^{i}=0.
\label{mag2}
\end{eqnarray}
where $\left( c^{i},\bar{c}^{i}\right) $, $\left(
c^{a},\bar{c}^{a}\right) $ stand for the diagonal and off-diagonal
Faddeev-Popov ghosts, while $\left( b^{i},b^{a}\right) $ denote the
diagonal and off-diagonal Lagrange multipliers. The covariant
derivative $D_{\mu }^{ab}$ in eq.(\ref{mag2}) is defined as
\begin{equation}
D_{\mu }^{ab}=\delta ^{ab}\partial _{\mu }-gf^{abi}A_{\mu }^{i}\;.
\label{mag3}
\end{equation}
Concerning the field strength $F_{\mu \nu }^{A}=(F_{\mu
\nu }^{i},F_{\mu \nu }^{a})$, we have
\begin{eqnarray}
F_{\mu \nu }^{a}\!\!\!\! &=&\!\!\!\!D_{\mu }^{ab}A_{\nu }^{b}-D_{\nu
}^{ab}A_{\mu }^{b}+gf^{abc}A_{\mu }^{b}A_{\nu }^{c}\;,  \label{mag4} \\
F_{\mu \nu }^{i}\!\!\!\! &=&\!\!\!\!\partial _{\mu }A_{\nu
}^{i}-\partial _{\nu }A_{\mu }^{i}+gf^{abi}A_{\mu }^{a}A_{\nu
}^{b}\;.  \nonumber
\end{eqnarray}
Thus, for the Yang-Mills action one has
\begin{equation}
S_{YM}=\frac{1}{4}\int d^{4}xF_{\mu \nu }^{A}F_{\mu \nu }^{A}=\frac{1}{4}%
\int d^{4}x\left( F_{\mu \nu }^{a}F_{\mu \nu }^{a}+F_{\mu \nu }^{i}F_{\mu
\nu }^{i}\right) \;.  \label{ym1}
\end{equation}\\\\ In order to interpolate between the Coulomb, the Landau and the maximal
Abelian gauges, we introduce the following, power counting
renormalizable,  gauge fixing term
\begin{equation}
S_{gf}=s_g\int{d^4}x\left[\bar{c}^a\left(a_{\mu\nu}\partial_\mu{A}_\nu^a+gK_{\mu\nu}f^{abi}A_\mu^bA_\nu^i+
\frac{\alpha}{2}b^a-\frac{\alpha}{2}gf^{abi}\bar{c}^bc^i-\frac{\beta}{4}gf^{abc}\bar{c}^bc^c\right)+
h_{\mu\nu}\bar{c}^i\partial_{\mu}A_\nu^i\right],\label{gf01}
\end{equation}
where $a_{\mu\nu}$, $h_{\mu\nu}$, $k_{\mu\nu}$ and $K_{\mu\nu}$ are
diagonal constant matrices defined as
\begin{eqnarray}
a_{\mu\nu}&\equiv&\mbox{diag}(1,1,1,a)\;,\nonumber\\
h_{\mu\nu}&\equiv&\mbox{diag}(1,1,1,h)\;,\nonumber\\
k_{\mu\nu}&\equiv&\mbox{diag}(k_s,k_s,k_s,k_t)\;,\nonumber\\
K_{\mu\nu}=\left(k_{\mu\sigma}-\delta_{\mu\sigma}\right)a_{\sigma\nu}&\equiv&
\mbox{diag}\left(k_s-1,k_s-1,k_s-1,a(k_t-1)\right)\;,\label{matrices}
\end{eqnarray}
and, $\left\{a,h,k_s,k_t,\alpha,\beta\right\}$ are gauge parameters.\\\\
The explicit expression for $S_{gf}$ reads
\begin{eqnarray}
S_{gf}&=&\int{d^4}x\left\{b^a\left[a_{\mu\nu}\partial_\mu{A}_\nu^a+
gK_{\mu\nu}f^{abi}A_\mu^bA_\nu^i+
\frac{\alpha}{2}b^a-\alpha{g}f^{abi}\bar{c}^bc^i-\frac{\beta}{2}gf^{abc}\bar{c}^bc^c\right]+
h_{\mu\nu}b^i\partial_{\mu}A_\nu^i\right.
\nonumber\\
&+&a_{\mu\nu}\bar{c}^a\partial_\mu{D}_\nu^{ab}c^b+
h_{\mu\nu}\bar{c}^i\partial_\mu\left(\partial_{\nu}c^i+gf^{abi}A_\nu^ac^b\right)-
ga_{\mu\nu}f^{abi}\partial_\mu\bar{c}^aA_\nu^bc^i+ga_{\mu\nu}f^{acd}\bar{c}^a\partial_\mu\left(A_\nu^cc^d\right)
\nonumber\\
&+&gK_{\mu\nu}f^{abi}A_\mu^a\partial_{\nu}c^i\bar{c}^b+
gK_{\mu\nu}f^{abi}\bar{c}^aA_\mu^iD_\nu^{bc}c^c+g^2K_{\mu\nu}f^{abi}f^{cdi}\bar{c}^ac^dA_\mu^bA_\nu^c
\nonumber\\
&+&g^2K_{\mu\nu}f^{abi}f^{bcj}A_\mu^iA_\nu^a\bar{c}^cc^j+g^2K_{\mu\nu}f^{abi}f^{bcd}A_\mu^iA_\nu^c\bar{c}^ac^d-
\frac{\alpha}{4}g^2f^{abi}f^{cdi}\bar{c}^a\bar{c}^bc^cc^d
\nonumber\\
&-&\left.\frac{\beta}{4}g^2f^{abc}f^{adi}\bar{c}^b\overline{c}^cc^dc^i-\frac{\beta}{8}g^2f^{abc}f^{ade}\bar{c}^b
\bar{c}^cc^dc^e\right\}\;.\label{gf0}
\end{eqnarray}
The gauge fixed action is then
\begin{equation}
S=S_{YM}+S_{gf}\;.\label{clm}
\end{equation}
where $S_{YM}$ and $S_{gf}$ are given by (\ref{ym1}) and
(\ref{gf0}), respectively.\\\\ The various gauges are attained in
the following way.
\begin{itemize}
\item The Landau gauge is achieved by setting
\begin{eqnarray}
a_{\mu\nu}\;=\;h_{\mu\nu}&=&\delta_{\mu\nu}\;,\nonumber\\
K_{\mu\nu}\;=\;\alpha\;=\;\beta&=&0\;.
\end{eqnarray}
namely
\begin{eqnarray}
a\;=\;h\;=k_t\;=\;k_s&=&1\;,\nonumber\\
\alpha\;=\;\beta&=&0 \label{l1} \;.
\end{eqnarray}
Substitution of the values (\ref{l1}) in the action (\ref{gf0})
provides the Landau gauge fixing,
\begin{equation}
S_{L}=s_g\int d^{4}x\left( \bar{c}^{a}\partial _{\mu }A_{\mu }^{a}+\bar{c}%
^{i}\partial _{\mu }A_{\mu }^{i}\right) =s_g\int d^{4}x\left( \bar{c}%
^{A}\partial _{\mu }A_{\mu }^{A}\right) \;,  \label{n10}
\end{equation}

\item The Coulomb gauge is achieved by setting
\begin{eqnarray}
a_{\mu\nu}\;=\;h_{\mu\nu}&\equiv&
\mbox{diag}(1,1,1,0)\;,\nonumber\\
K_{\mu\nu}\;=\;\alpha\;=\;\beta&=&0\;,
\end{eqnarray}
{\it i.e.}
\begin{eqnarray}
a\;=\;h\;=\;\alpha\;=\;\beta&=&0\nonumber\\
k_t\;=\;k_s&=&1 \label{c1} \;.
\end{eqnarray}
The substitution of (\ref{c1}) in the action (\ref{gf0}) gives
the Coulomb gauge fixing,
\begin{equation}
S_{C}=s_g\int{d^4}x\left(\bar{c}^a\partial_kA_k^a+\bar{c}^i\partial_kA_k^i\right)=
s_g\int{d^4}x\;\bar{c}^A\partial_kA_k^A\;. \label{n8}
\end{equation}
\item The Maximal Abelian gauge is achieved by setting
\begin{eqnarray}
a_{\mu\nu}\;=\;h_{\mu\nu}\;=\;K_{\mu\nu}&=&\delta_{\mu\nu}\;,\nonumber\\
\beta&=&\alpha\;,
\end{eqnarray}
{\it i.e.}
\begin{eqnarray}
a\;=\;h&=&1\;,\nonumber\\
k_t\;=\;k_s&=&0\;,\nonumber\\
\beta&=&\alpha\;.
\end{eqnarray}
Therefore, for the Maximal Abelian gauge fixing\footnote{As is
known, the gauge parameter $\alpha$ has to be introduced for
renormalization purposes. The real MAG condition, namely
$D_\mu^{ab}A_\mu^b =0$, is attained in the limit
$\alpha\rightarrow 0$, which has to be taken after renormalization
\cite{Dudal:2004rx}.} we get
\begin{equation}
S_{MAG}=s_g\int{d^4}x\left[\bar{c}^a\left(D_\mu^{ab}A_\mu^b+\frac{\alpha}{2}b^a-
\frac{\alpha}{2}gf^{abi}\bar{c}^bc^i-\frac{\alpha}{4}gf^{abc}\bar{c}^bc^c\right)+
\bar{c}^i\partial_\mu{A}_\mu^i\right]\;. \label{n12}
\end{equation}

\end{itemize}

\noindent We point out that the gauge fixing (\ref{gf0}) is slightly
more general that the one reported in \cite{Capri:2005zj}, which was
in fact limited only to tree level aspects. As it will be shown in
the rest of this article, the gauge fixing (\ref{gf0}) turns out to
be suitable in order to
establish the multiplicative renormalizability of the model. \\\\
Let us now proceed by discussing the Lorentz breaking induced by
the gauge (\ref{gf0}) and the way to control it.

\section{BRST quantization and the breaking of Lorentz invariance}

A problem to be faced in Yang-Mills theories quantized in the
gauge (\ref{gf0}) is that of the  breaking of the Lorentz
invariance. This problem was successfully treated in
\cite{Baulieu:1998kx} for the case of the Coulomb-Landau
interpolating gauge. Here, we will use the same technique in order
to control the Lorentz breaking. We refer thus to
\cite{Baulieu:1998kx} and references therein for all details. This
section is then devoted to the treatment of the Lorentz breaking
in the specific case of the CLM gauges.

\subsection{BRST quantization method}

Let us start by recalling the main steps of the BRST quantization
method \cite{Becchi:1975nq,Tyutin:1975qk}. For that we consider a
general gauge model with classical action $S(A)$ and coupling $g$,
where $A$ is the gauge field. The action $S(A)$ is invariant under
transformations of a certain Lie group $G$ with elements
\begin{equation}
u=e^{\omega}\;\in\;G\;|\;\omega=\omega^\mathcal{A}\lambda^\mathcal{A}\;,
\end{equation}
where $\lambda^\mathcal{A}$ are the group generators. The index
$\mathcal{A}$ is used here as a general index, with arbitrary
dimension. The algebra of the generators is, typically, given by
\begin{equation}
[\lambda^\mathcal{A},\lambda^\mathcal{B}]=f^{\mathcal{A}\mathcal{B}\mathcal{C}}\lambda^\mathcal{C}\;,\label{alg1}
\end{equation}
where $f^{\mathcal{A}\mathcal{B}\mathcal{C}}$ are the structure
constants of the group. The BRST method of quantization amounts to
introduce a set of Lie algebra valued anticommuting fields,
$C=C^\mathcal{A}\lambda^\mathcal{A}$, for each generator of the
symmetry, the so called ghost fields or Faddeev-Popov ghosts.
Together with the ghost fields, a set of nilpotent transformations
is obtained, giving rise to the BRST transformations. For the
transformation of the ghost fields we have\footnote{Notice that
the appearance of the coupling constant $g$ in eq.(\ref{brs1}) is
just a matter of convention. Here we adopt a different convention
than that employed in \cite{Baulieu:1998kx}.}
\begin{equation}
sC^\mathcal{A}=\frac{g}{2}f^{\mathcal{A}\mathcal{B}\mathcal{C}}C^\mathcal{B}C^\mathcal{C}\;,\label{brs1}
\end{equation}
which is just the Maurer-Cartan structure equation of the Lie
group. The operator $s$ is the BRST operator. For the gauge field
the BRST transformation is, in fact, obtained by replacing the
group element parameter $\omega$ by the ghost field $C$, namely
\begin{equation}
s A^\mathcal{A}_{\mu} = -(\partial_{\mu} C^\mathcal{A} +
gf^{\mathcal{A}\mathcal{B}\mathcal{C}}A^\mathcal{B}_{\mu}
C^\mathcal{C} )\;, \label{brsa1}
\end{equation}
ensuring thus the nilpotency of the BRST operator,
$s^2=0$.\\\\From the gauge invariance of the action $S(A)$, it
follows that
\begin{equation}
s S(A) = 0 \;. \label{brsa2}
\end{equation}
The quantization of the theory is achieved by introducing a gauge
fixing in a BRST invariant way. Suppose that the constraint
expressing the gauge condition is given by the equation
\begin{equation}
f^\mathcal{A}(A)=0\;.\label{const1}
\end{equation}
The gauge fixed action then reads
\begin{equation}
S=S(A)+s\Delta^{-1}_{gf}(f,C,\bar{C},b)\;,
\end{equation}
where $\Delta^{-1}_{gf}$ is a local polynomial with ghost number
$-1$ and dimension four,
\begin{equation}
\Delta^{-1}_{gf}(f,C,\bar{C},b)=\int{d^4x}\left\{\bar{C}^\mathcal{A}\left[f^\mathcal{A}(A)+
\frac{\alpha}{2}b^\mathcal{A}\right]+\delta^{-1}(\bar{C},C)\right\}\;.\label{const2}
\end{equation}
Here $b$ and $\bar C$ are respectively the Lautrup-Nakanishi and
anti-ghost fields, transforming as a BRST doublet, {\it i.e.}
\begin{eqnarray}
s\bar{C}&=&b\;,\nonumber\\
s{b}&=&0\;.\label{brs2}
\end{eqnarray}
The parameter $\alpha$ is a gauge parameter, and the polynomial
function $\delta^{-1}$ might be necessary for renormalizability
purposes. For example, nonlinear gauges need a quartic ghost
self-interaction generated, for instance, by
\begin{equation}
\delta^{-1} = f^{\mathcal{A}\mathcal{B}\mathcal{C}} {\bar
C}^{\mathcal{A}} C^\mathcal{B} C^\mathcal{C} \;. \label{qua1}
\end{equation}

\subsection{Controlling the Lorentz breaking}

Now we apply the BRST quantization method to our gauge fixing
(\ref{gf0}). First, one has to identify the symmetries broken by the
gauge fixing term. In our case these are: the local $SU(N)$ gauge
symmetry, the global color invariance due to the Abelian
decomposition, eq.(\ref{mag1}),  and the Lorentz symmetry. Here, the
Lorentz symmetry coincides in fact with the rotation group $O(4)$,
because we are dealing with a four dimensional Euclidean space-time.
Thus, we shall make use of a BRST symmetry associated to the large
symmetry group $SU(N)\otimes{O}(4)$, this will give rise to the
introduction of a set of ghost fields which will enable us to
control the gauge breaking together with the Lorentz breaking in a
powerful and simple fashion. The algebra obeyed by the generators of
this group is
\begin{eqnarray}
\left[T^a,T^b\right]&=&f^{abc}T^c+f^{abi}T^i\;,\nonumber\\
\left[T^a,T^i\right]&=&f^{abi}T^b\;,\nonumber\\
\left[T^i,T^j\right]&=&0\;,\nonumber\\
\left[\Sigma_{\mu\nu},\Sigma_{\gamma\delta}\right]&=&f_{\mu\nu\gamma\delta\alpha\beta}
\Sigma_{\alpha\beta}\;,\nonumber\\
\left[\Sigma_{\mu\nu},T^a\right]&=&f^{ab}_{\mu\nu}T^b\;,\nonumber\\
\left[\Sigma_{\mu\nu},T^i\right]&=&f^{ij}_{\mu\nu}T^j\;,\label{alg2}
\end{eqnarray}
where $T^a$ and $T^i$ are, respectively, the generators of the
non-Abelian and Abelian part of the gauge group $SU(N)$, and
$\Sigma_{\mu\nu}$ are the generators of the rotation group $O(4)$.
As already explained in Section 2, the indices $(a,b,c)$ refer to
the $N(N-1)$ off-diagonal generators, while the indices $(i,j)$
label the $(N-1)$ diagonal generators of the Cartan subgroup of
$SU(N)$. The structure constants of $O(4)$ are given by
\begin{equation}
f_{\mu\nu\rho\sigma\alpha\beta}=-\frac{1}{2}\left[\left(\delta_{\mu\sigma}\delta_{\alpha\nu}-
\delta_{\nu\sigma}\delta_{\alpha\mu}\right)\delta_{\rho\beta}+\left(\delta_{\nu\rho}\delta_{\mu\alpha}-
\delta_{\mu\rho}\delta_{\alpha\nu}\right)\delta_{\sigma\beta}\right]\;,\label{strucconst}
\end{equation}
while the mixed structure constants are found to be
\begin{eqnarray}
f^{ab}_{\mu\nu}&=&-\frac{1}{2}\delta^{ab}\left(x_\mu\partial_\nu-x_\nu\partial_\mu\right)\;,\nonumber\\
f^{ij}_{\mu\nu}&=&-\frac{1}{2}\delta^{ij}\left(x_\mu\partial_\nu-x_\nu\partial_\mu\right)\;.
\end{eqnarray} \\\\In addition of the Faddeev-Popov ghost,
$\left\{c^a,c^i\right\}$, corresponding to the gauge symmetry
$SU(N)$ we have to include further global ghosts,
$\{V_{\mu\nu}\}$,  associated with the global $O(4)$ symmetry
breaking. These global ghosts, being
space-time independent, will behave as external sources.\\\\
According to the group structure (\ref{alg2}) and to equation
(\ref{brs1}),  it is straightforward to deduce the BRST
transformations of all ghosts
\begin{eqnarray}
sc^a&=&gf^{abi}c^bc^i+\frac{g}{2}f^{abc}c^bc^c-gV_{\mu\nu}x_\mu\partial_\nu{c}^a\;,\nonumber\\
sc^i&=&\frac{g}{2}f^{abi}c^ac^b-gV_{\mu\nu}x_\mu\partial_\nu{c}^i\;,\nonumber\\
sV_{\mu\nu}&=&-gV_{\mu\gamma}V_{\gamma\nu}\;.\label{brs3}
\end{eqnarray}
Also, for the BRST doublets one has
\begin{eqnarray}
s\bar{c}^a&=&b^a\;,\nonumber\\
s\bar{c}^i&=&b^i\;,\nonumber\\
sb^a&=&0\;,\nonumber\\
sb^i&=&0\;.\label{brs4}
\end{eqnarray}
The BRST transformations of the gauge fields are those
corresponding to infinitesimal gauge transformations of the large
group $SU(N)\otimes{O}(4)$, maintaining the nilpotence of $s$,
\begin{eqnarray}
sA_\mu^a&=&-D^{ab}_\mu{c}^b-gf^{abc}A_\mu^bc^c-gf^{abi}A_\mu^bc^i-gV_{\mu\nu}A_\nu^a-gV_{\gamma\nu}x_\gamma\partial_\nu
A_\mu^a\;,\nonumber\\
sA_\mu^i&=&-\partial_\mu{c}^i-gf^{abi}A_\mu^ac^b-gV_{\mu\nu}A_\nu^i-gV_{\gamma\nu}x_\gamma\partial_\nu{A}_\mu^i\;.
\label{brs5}
\end{eqnarray}
Such extended BRST symmetry (\ref{brs3}-\ref{brs5}) encodes the
Lorentz rotations. As such, it is a generalization of the previous one, eq.(\ref{mag2}).\\\\
Notice that the extended BRST operator $s$ can be decomposed into
an ordinary BRST operator $s_g$ and a Lorentz BRST operator $s_L$,
\begin{equation}
s=s_g+s_L\;,
\end{equation}
obeying
\begin{equation}
s^2=s_g^2=s_L^2=\left\{s_g,s_L\right\}=0\;.
\end{equation}
The operator $s_g$ corresponds to the pure gauge sector and is
given in eq.(\ref{mag2}), together with
\begin{equation}
s_gV_{\mu\nu}=0\;.
\end{equation}
The Lorentz BRST operator acts on the fields like
\begin{eqnarray}
s_LA_\mu^{a,i}&=&-gV_{\mu\nu}A_\nu^{a,i}-gV_{\alpha\beta}x_\alpha\partial_\beta{A}_\mu^{a,i}\;,\nonumber\\
s_Lc^{a,i}&=&-gV_{\alpha\beta}x_\alpha\partial_\beta{c}^{a,i}\;,\nonumber\\
s_L\bar{c}^{a,i}&=&0\;,\nonumber\\
s_Lb^{a,i}&=&0\;,\nonumber\\
s_LV_{\mu\nu}&=&-gV_{\mu\gamma}V_{\gamma\nu}\;.
\end{eqnarray}

\section{Renormalizability of the CLM gauges}

Having defined the structure of the BRST operator, we can write
down the gauge fixing term which has to be added to the Yang-Mills
action (\ref{ym1}). This is done by replacing the ordinary BRST
operator $s_g$, in expression (\ref{gf01}), by the extended BRST
operator $s$ defined in (\ref{brs3}-\ref{brs5}). Thus,
\begin{equation}
S_{gf}=s\int{d^4}x\left[\bar{c}^a\left(a_{\mu\nu}\partial_\mu{A}_\nu^a+gK_{\mu\nu}f^{abi}A_\mu^bA_\nu^i+
\frac{\alpha}{2}b^a-\frac{\alpha}{2}gf^{abi}\bar{c}^bc^i-\frac{\beta}{4}gf^{abc}\bar{c}^bc^c\right)+
h_{\mu\nu}\bar{c}^i\partial_{\mu}A_\nu^i\right],\label{gf1}
\end{equation}
yielding
\begin{eqnarray}
S_{gf}&=&\int{d^4}x\left\{b^a\left[a_{\mu\nu}\partial_\mu{A}_\nu^a+
gK_{\mu\nu}f^{abi}A_\mu^bA_\nu^i+
\frac{\alpha}{2}b^a-\alpha{g}f^{abi}\bar{c}^bc^i-\frac{\beta}{2}gf^{abc}\bar{c}^bc^c\right]+
h_{\mu\nu}b^i\partial_{\mu}A_\nu^i\right.
\nonumber\\
&+&a_{\mu\nu}\bar{c}^a\partial_\mu{D}_\nu^{ab}c^b+
h_{\mu\nu}\bar{c}^i\partial_\mu\left(\partial_{\nu}c^i+gf^{abi}A_\nu^ac^b\right)-
ga_{\mu\nu}f^{abi}\partial_\mu\bar{c}^aA_\nu^bc^i+ga_{\mu\nu}f^{acd}\bar{c}^a\partial_\mu\left(A_\nu^cc^d\right)
\nonumber\\
&+&gK_{\mu\nu}f^{abi}A_\mu^a\partial_{\nu}c^i\bar{c}^b+
gK_{\mu\nu}f^{abi}\bar{c}^aA_\mu^iD_\nu^{bc}c^c+g^2K_{\mu\nu}f^{abi}f^{cdi}\bar{c}^ac^dA_\mu^bA_\nu^c
\nonumber\\
&+&g^2K_{\mu\nu}f^{abi}f^{bcj}A_\mu^iA_\nu^a\bar{c}^cc^j+g^2K_{\mu\nu}f^{abi}f^{bcd}A_\mu^iA_\nu^c\bar{c}^ac^d-
\frac{\alpha}{4}g^2f^{abi}f^{cdi}\bar{c}^a\bar{c}^bc^cc^d
\nonumber\\
&-&\left.\frac{\beta}{4}g^2f^{abc}f^{adi}\bar{c}^b\overline{c}^cc^dc^i-\frac{\beta}{8}g^2f^{abc}f^{ade}\bar{c}^b
\bar{c}^cc^dc^e\right\}+g\int{d^4x}\left\{V_{\mu\sigma}\left(a_{\mu\nu}\partial_\nu\bar{c}^aA_\sigma^a+
h_{\mu\nu}\partial_\nu\bar{c}^iA_\sigma^i\right)\right.
\nonumber\\
&+&\left.V_{\alpha\beta}x_\alpha\left(a_{\mu\nu}\partial_\beta{A}_\mu^a\partial_\nu\bar{c}^a+
h_{\mu\nu}\partial_\beta{A}_\mu^i\partial_\nu\bar{c}^i\right)+
gK_{\mu\sigma}f^{abi}\bar{c}^aV_{\mu\nu}\left(A_\sigma^iA_\nu^b+A_\sigma^bA_\nu^i\right)\right.
\nonumber\\
&+&\left.gK_{\mu\nu}f^{abi}V_{\alpha\beta}x_\alpha\partial_\beta\bar{c}^aA_\mu^bA_\nu^i-
\alpha{g}f^{abi}V_{\mu\nu}x_\mu\partial_\nu\bar{c}^a\bar{c}^bc^i-
\frac{\beta}{2}gf^{abc}V_{\mu\nu}x_\mu\partial_\nu\bar{c}^a\bar{c}^bc^c\right\}\;.\label{gf}
\end{eqnarray}

\subsection{Ward identities}

In order to establish the set of Ward identities describing the
symmetries displayed by the gauge fixed action (\ref{clm}), one
has to introduce external sources $(\Omega, L, M)$ coupled to the
nonlinear BRST transformations \cite{book}. Thus, the complete
action we will work with is
\begin{equation}
\Sigma=S_{YM}+S_{gf}+S_{ext}\;,\label{qclm}
\end{equation}
where
\begin{eqnarray}
S_{ext}&=&s\left[\int{d^4x}\left(-\Omega_\mu^aA_\mu^a-\Omega_\mu^iA_\mu^i+L^ac^a+L^ic^i\right)+M_{\mu\nu}V_{\mu\nu}\right]\nonumber\\
&=&\int{d^4x}\left\{-\Omega_\mu^a\left(D_\mu^{ab}c^b+gf^{abi}A_\mu^bc^i+gf^{abc}A_\mu^bc^c+gV_{\mu\nu}A_\nu^a+
gV_{\alpha\beta}x_\alpha\partial_\beta{A}_\mu^a\right)-\Omega_\mu^i\left(\partial_\mu{c}^i\right.\right.\nonumber\\
&+&\left.\left.gf^{abi}A_\mu^ac^b+gV_{\mu\nu}A_\nu^i+gV_{\alpha\beta}x_\alpha\partial_\beta{A}_\mu^i\right)+
gL^a\left(f^{abi}c^bc^i+\frac{1}{2}f^{abc}c^bc^c-V_{\mu\nu}x_\mu\partial_\nu{c}^a\right)\right.\nonumber\\
&+&\left.gL^i\left(\frac{1}{2}f^{abi}c^ac^b-V_{\mu\nu}x_\mu\partial_\nu{c}^i\right)\right\}-
gM_{\mu\nu}V_{\mu\gamma}V_{\gamma\nu}\;.\label{ext}
\end{eqnarray}
Notice that $V$ is a constant field, and so is $M$. The total set
of fields and sources, with their corresponding ghost numbers and
dimensions, is displayed in Table \ref{table1}.

\begin{table}[t]
\centering
\begin{tabular}{|c|c|c|c|c|c|c|c|c|}
\hline
fields / sources & $A$ & $V$ & $c$ & $\bar{c}$ & $b$ & $\Omega$ & $L$ & $M$ \\
\hline
dimension & 1 & 0 & 0 & 2 & 2 & 3 & 4 & 4 \\
ghost number & 0 & 1 & 1 & $-1$ & 0 & $-1$ & $-2$ & $-2$ \\
\hline
\end{tabular}
\caption{Dimension and ghost number of the fields and sources.}
\label{table1}
\end{table}

\noindent The complete action (\ref{qclm}) obeys the following set
of Ward identities:
\begin{itemize}
\item The Slavnov-Taylor identity
\begin{equation}
\mathcal{S}(\Sigma)=0 \;,\label{st0}
\end{equation}
with
\begin{eqnarray}
\mathcal{S}(\Sigma)&=&\int{d^4x}\left(\frac{\delta\Sigma}{\delta\Omega^a_\mu}\frac{\delta\Sigma}{\delta{A}^a_\mu}+
\frac{\delta\Sigma}{\delta\Omega^i_\mu}\frac{\delta\Sigma}{\delta{A}^i_\mu}+
\frac{\delta\Sigma}{\delta{L}^a}\frac{\delta\Sigma}{\delta{c}^a}+
\frac{\delta\Sigma}{\delta{L}^i}\frac{\delta\Sigma}{\delta{c}^i}
+b^a\frac{\delta\Sigma}{\delta\bar{c}^a}+b^i\frac{\delta\Sigma}{\delta\bar{c}^i}\right)\nonumber\\
 &+&\frac{\delta{\Sigma}}{\delta{M_{\mu\nu}}}\frac{\delta{\Sigma}}{\delta{V}_{\mu\nu}}\;.\label{st}
\end{eqnarray}

\item The diagonal gauge fixing and diagonal antighost equations
\begin{eqnarray}
\frac{\delta\Sigma}{\delta{b}^i}&=&h_{\mu\nu}\partial_\mu{A}_\nu^i\;,\nonumber\\
\frac{\delta\Sigma}{\delta\bar{c}^i}+h_{\mu\nu}\partial_\mu\frac{\delta\Sigma}{\delta\Omega_\nu^i}&=&0\;.
\label{antighost}
\end{eqnarray}

\item The linearly broken integrated diagonal ghost Ward identity
\begin{equation}
\int{d^4x}\left(\frac{\delta\Sigma}{\delta{c}^i}+gf^{abi}\bar{c}^a\frac{\delta\Sigma}{\delta{b}^b}\right)=
gf^{abi}\int{d^4x}\left(\Omega_\mu^aA_\mu^b-L^ac^b\right)\;.\label{ghost}
\end{equation}

\item The $M$ equation
\begin{equation}
\frac{\delta\Sigma}{\delta{M}_{\mu\nu}}=-gV_{\mu\gamma}V_{\gamma\nu}\;.\label{m}
\end{equation}
\end{itemize}
The last Ward identity is possible only because $V$ is a global
ghost field. In addition, we also have a residual global
$U(1)^{N-1}$ symmetry and a residual global $O(3)$ symmetry,
corresponding to the three-space rotations. As shown in
\cite{Fazio:2001rm}, the $U(1)^{N-1}$ global symmetry follows by
anticommuting the diagonal ghost equation (\ref{ghost}) with the
Slavnov-Taylor identities (\ref{st}). The $O(3)$ symmetry will be
tacitly assumed in the construction of the counterterm.

\subsection{General invariant counterterm}

Let us face now the construction of the most general counterterm
consistent with the symmetries of the action (\ref{qclm}).
Following the algebraic renormalization theory \cite{book}, this
can be performed by adding a generic local field functional,
$\Sigma^c$, to the classical action (\ref{qclm})
\begin{equation}
{\widetilde \Sigma}=\Sigma+\epsilon\Sigma^c\;,
\end{equation}
where $\epsilon$ is a small expansion parameter. Imposing now the
Ward identities (\ref{st0},\ref{antighost}-\ref{m}), one obtains
that $\Sigma^c$ has to fulfill the following set of constraints
\begin{eqnarray}
\mathcal{B}_\Sigma\Sigma^c&=&0\;,\nonumber\\
\frac{\delta\Sigma^c}{\delta{b}^i}&=&0\;,\nonumber\\
\frac{\delta\Sigma^c}{\delta\bar{c}^i}+h_{\mu\nu}\partial_\mu\frac{\delta\Sigma^c}{\delta\Omega_\nu^i}&=&0\;,\nonumber\\
\int{d^4x}\left(\frac{\delta\Sigma^c}{\delta{c}^i}+gf^{abi}\bar{c}^b\frac{\delta\Sigma^c}{\delta{b}^c}\right)&=&0\;,
\nonumber\\
\frac{\delta\Sigma^c}{\delta{M}_{\mu\nu}}&=&0\;,\label{cons}
\end{eqnarray}
where $\mathcal{B}_\Sigma$ is the nilpotent, $\mathcal{B}_\Sigma
\mathcal{B}_\Sigma=0$,  linearized Slavnov-Taylor operator,
\begin{eqnarray}
\mathcal{B}_\Sigma&\equiv&\int{d^4x}\left(\frac{\delta\Sigma}{\delta\Omega^a_\mu}\frac{\delta}{\delta{A}^a_\mu}+
\frac{\delta\Sigma}{\delta{A}^a_\mu}\frac{\delta}{\delta\Omega^a_\mu}+
\frac{\delta\Sigma}{\delta\Omega^i_\mu}\frac{\delta}{\delta{A}^i_\mu}+
\frac{\delta\Sigma}{\delta{A}^i_\mu}\frac{\delta}{\delta\Omega^i_\mu}+
\frac{\delta\Sigma}{\delta{L}^a}\frac{\delta}{\delta{c}^a}+
\right.\nonumber\\
&&+\left.\frac{\delta\Sigma}{\delta{c}^a}\frac{\delta}{\delta{L}^a}
+\frac{\delta\Sigma}{\delta{L}^i}\frac{\delta}{\delta{c}^i}+
\frac{\delta\Sigma}{\delta{c}^i}\frac{\delta}{\delta{L}^i}+
b^a\frac{\delta}{\delta\bar{c}^a}+b^i\frac{\delta\Sigma}{\delta\bar{c}^i}\right)\nonumber\\
&&+\frac{\delta{\Sigma}}{\delta{M_{\mu\nu}}}\frac{\delta}{\delta{V}_{\mu\nu}}+
\frac{\delta{\Sigma}}{\delta{V}_{\mu\nu}}\frac{\delta}{\delta{M_{\mu\nu}}}\;.\label{lst}
\end{eqnarray}
From general cohomological arguments \cite{book}, it follows that
the first condition of eqs.(\ref{cons}) implies that $\Sigma^c$
can be written as
\begin{equation}
\Sigma^c=a_0S_{YM}+\mathcal{B}_\Sigma\Delta^{-1}\;,\label{count}
\end{equation}
where $\Delta^{-1}$ is the most general local polynomial in the
fields with dimension 4 and ghost number -1. Furthermore, from the
constraints (\ref{cons}), it  turns out that
\begin{eqnarray}
\Delta^{-1}&=&\int{d^4x}\left\{a_1\alpha\bar{c}^a\left(b^a-gf^{abi}\bar{c}^bc^i\right)+
a_5\beta{g}f^{abc}\bar{c}^a\bar{c}^bc^c+a_3L^ac^a
\right.\nonumber\\
&+&\left.
a_{\mu\nu}d_{\nu\sigma}^{(8,9)}\bar{c}^a\partial_\mu{A}_\sigma^a
+a_{\mu\nu}\left(k_{\nu\sigma}d_{\sigma\gamma}^{(12,13)}-d_{\nu\gamma}^{(8,9)}\right)
gf^{abi}\bar{c}^aA_\mu^bA_\gamma^i
\right.\nonumber\\
&+&\left.d_{\mu\nu}^{(14,15)}\Omega_\mu^aA_\nu^a
-d_{\mu\sigma}^{(10,11)}\left(h_{\mu\nu}\partial_\nu\bar{c}^i+\Omega_\mu^i\right)A_\sigma^i\right\}\;,
 \label{deltaminus0}
\end{eqnarray}
where
\begin{equation}
d_{\mu\nu}^{(i,j)}\equiv \mbox{diag}\left(a_j,a_j,a_j,a_i
\right)\;.
\end{equation}
Explicitly, (\ref{deltaminus0}) reads
\begin{eqnarray}
\Delta^{-1}&=&\int{d^4x}\left\{a_1\alpha\bar{c}^a\left(b^a-gf^{abi}\bar{c}^bc^i\right)+
a_5\beta{g}f^{abc}\bar{c}^a\bar{c}^bc^c+a_3L^ac^a+a_8a\bar{c}^aD_4^{ab}A_4^b\right.\nonumber\\
&+&a_9\bar{c}^aD_k^{ab}A_k^b+
a_{12}k_tagf^{abi}\bar{c}^aA_4^bA_4^i+a_{13}k_sgf^{abi}\bar{c}^aA_k^bA_k^i+a_{14}\Omega_4^aA_4^a+a_{15}\Omega_k^aA_k^a
\nonumber\\
&-&\left.a_{10}\left(h\partial_4\bar{c}^i+\Omega_4^i\right)A_4^i-a_{11}\left(\partial_k\bar{c}^i+
\Omega_k^i\right)A_k^i\right\}\;.\label{deltaminus1}
\end{eqnarray}
The parameters
$(a_0,a_1,a_3,a_5,a_8,a_9,a_{10},a_{11},a_{12},a_{13},a_{14},a_{15})$
are free independent constants. Thus, in order to establish the
renormalizability of our model, we have to show that these twelve
independent parameters can be reabsorbed in the action, through a
renormalization of the fields, parameters and external sources.
This will be the task of the next section.

\subsection{Stability}

It remains to show that the counterterm (\ref{count}) can be in
fact reabsorbed by means of a multiplicative renormalization of
the fields and parameters, according to
\begin{equation}
\Sigma(\phi_0,J_0,\xi_0)=\Sigma(\phi,J,\xi)+\epsilon\Sigma^c(\phi,J,\xi)
+ O(\epsilon^2) \;,\label{ren1}
\end{equation}
where
\begin{eqnarray}
\phi^a_0&=&Z^{1/2}_\phi\phi^a\;,\nonumber\\
\phi^i_0&=&z^{1/2}_\phi\phi^i\;,\nonumber\\
J^a&=&Z_JJ^a\;,\nonumber\\
J^i&=&z_JJ^i\;,\nonumber\\
\xi_0&=&Z_\xi\xi\;,\label{ren2}
\end{eqnarray}
and
\begin{eqnarray}
\phi&\in&\{A,c,\bar{c},V\}\;,\nonumber\\
J&\in&\{\Omega,L,M\}\;,\nonumber\\
\xi&\in&\{g,\alpha,\beta,a,h,k_t,k_s\}\;.\label{ren3}
\end{eqnarray}
Due to the use of a non covariant gauge fixing, we shall
distinguish the renormalization of the fourth component of the
gluon field from the remaining three components, according to
\begin{eqnarray}
A^a_{04}&=&\tilde{Z}^{1/2}_AA^a_4\;,\nonumber\\
A^a_{0k}&=&Z^{1/2}_AA^a_k\;,\nonumber\\
A^i_{04}&=&\tilde{z}^{1/2}_AA^a_4\;,\nonumber\\
A^i_{0k}&=&z^{1/2}_AA^a_k\;.\label{example}
\end{eqnarray}\\\\ The corresponding renormalization factors are given by
\begin{eqnarray}
\tilde{Z}^{1/2}_A&=&1+\epsilon\left(\frac{a_0}{2}+a_{14}\right)\;,\nonumber\\
Z^{1/2}_A&=&1+\epsilon\left(\frac{a_0}{2}+a_{15}\right)\;,\nonumber\\
\tilde{z}^{1/2}_A&=&1+\epsilon\left(\frac{a_0}{2}-a_{10}\right)\;,\nonumber\\
z^{1/2}_A&=&1+\epsilon\left(\frac{a_0}{2}-a_{11}\right)\;.\label{rengluon}
\end{eqnarray}
For the renormalization of the coupling constant one has
\begin{equation}
Z_{g}=1-\epsilon\frac{a_0}{2}\;.\label{reng}
\end{equation}
For the space-time gauge parameters $a$ and $h$ the renormalization
factors reads
\begin{eqnarray}
Z_a&=&1+\epsilon\left(a_8-a_9\right)\;,\nonumber\\
Z_h&=&z_A^{1/2}\tilde{z}_A^{-1/2}\;,\nonumber\\
\label{renah}
\end{eqnarray}
while for the gauge group parameters we have
\begin{eqnarray}
Z_\alpha&=&1+\epsilon\left(a_0+2a_1-2a_9\right)\;,\nonumber\\
Z_\beta&=&1+\epsilon\left(a_0-4a_5-2a_9\right)\;,\nonumber\\
Z_{k_t}&=&1+\epsilon\left(-a_8+a_{12}\right)\;,\nonumber\\
Z_{k_s}&=&1+\epsilon\left(-a_9+a_{13}\right)\;.\label{renak}
\end{eqnarray}
The Faddeev-Popov ghosts renormalize according to
\begin{eqnarray}
Z_c^{1/2}&=&1+\epsilon\left(-a_3+\frac{a_{11}}{2}\right)\;,\nonumber\\
Z_{\bar{c}}^{1/2}&=&1+\epsilon\left(a_9-\frac{a_{11}}{2}\right)\;,\nonumber\\
z_c^{1/2}&=&z_{\bar{c}}^{1/2}\;=\;1+\epsilon\frac{a_{11}}{2}\;=\;Z_g^{-1/2}z_A^{-1/4}\;.\label{renC}
\end{eqnarray}
For the Lagrange multipliers we obtain
\begin{eqnarray}
Z_b^{1/2}&=&1+\epsilon\left(-\frac{a_0}{2}+a_9\right)\;=\;Z_g^{1/2}z_A^{-1/4}Z_{\bar{c}}^{1/2}\;,\nonumber\\
z_b^{1/2}&=&z_A^{-1/2}\;,\label{renb}
\end{eqnarray}
and, for the external sources $\Omega$ and $L$
\begin{eqnarray}
\tilde{Z}_\Omega&=&1-\epsilon\left(\frac{a_{11}}{2}+a_{14}\right)\;=\;Z_g^{-1/2}z_A^{1/4}\tilde{Z}_A^{-1/2}\;,
\nonumber\\
Z_\Omega&=&1-\epsilon\left(\frac{a_{11}}{2}+a_{15}\right)\;=\;Z_g^{-1/2}z_A^{1/4}Z_A^{-1/2}\;,\nonumber\\
\tilde{z}_\Omega&=&1+\epsilon\left(a_{10}-\frac{a_{11}}{2}\right)\;=\;Z_g^{-1/2}z_A^{1/4}\tilde{z}_A^{-1/2}\;,\nonumber\\
z_\Omega&=&1+\epsilon\frac{a_{11}}{2}\;=\;Z_g^{-1/2}z_A^{-1/4}\;,\nonumber\\
Z_L&=&Z_g^{-1}Z_c^{-1/2}z_c^{-1/2}\;,\nonumber\\
z_L&=&Z_g^{-1}z_c^{-1}\;.\label{renJ}
\end{eqnarray}
Finally, the Lorentz ghost and its associated BRST external source
renormalize as
\begin{eqnarray}
Z_V^{1/2}&=&z_c^{1/2}\;,\nonumber\\
Z_M&=&Z_g^{-1}z_c^{-1}\;.\label{renVM}
\end{eqnarray}\\\\ This ends the proof of the multiplicatively renormalizability of
Yang-Mills theory quantized in the general interpolating gauge
(\ref{gf0}).

\section{Conclusions}

In this work we have proven the renormalizability of a generalized
gauge fixing which interpolates between the Coulomb, the Landau
and the maximal Abelian gauges.\\\\ It is worth underlining that
all these three gauges are extensively used in lattice numerical
simulations. The introduction of such a generalized interpolating
gauges seems thus appropriate, as it could be helpful in order to
achieve a kind of unifying understanding of the physical operators
in all these gauges.

\section*{Acknowledgments}

We thank the Conselho Nacional de Desenvolvimento Cient\'{\i}fico
e Tecnol\'{o}gico (CNPq-Brazil), the Faperj, Funda{\c c}{\~a}o de
Amparo {\`a} Pesquisa do Estado do Rio de Janeiro, the SR2-UERJ
and the Coordena{\c{c}}{\~{a}}o de Aperfei{\c{c}}oamento de
Pessoal de N{\'\i}vel Superior (CAPES) for financial support.

\appendix

\section{Minimizing functional}\label{apA}

In this Appendix we discuss the possibility of introducing a
suitable minimizing functional for the interpolating gauge, a
feature which could allow for a lattice implementation of this
generalized gauge fixing. For that, we shall consider the
interpolating gauge (\ref{gf0}) when
\begin{eqnarray}
h_{\mu\nu}&\rightarrow&a_{\mu\nu}\;,\nonumber\\
k_t&\rightarrow&k\;,\nonumber\\
k_s&\rightarrow&k\;,\nonumber\\
\beta&\rightarrow&\alpha\;,\label{limit0}
\end{eqnarray}
which gives
\begin{eqnarray}
S_{class}&=&\int{d^4}x\left\{b^{a}\left[a_{\mu\nu}\partial_\mu^aA_\nu^a+
(k-1)gf^{abi}a_{\mu\nu}A_\mu^iA_\nu^b+\frac{\alpha}{2}b^a-\alpha{g}f^{abi}\bar{c}^bc^i-
\frac{\alpha}{2}gf^{abc}\bar{c}^bc^c\right]\right.\nonumber\\
&+&b^ia_{\mu\nu}\partial_\mu{A}_\nu^i+\bar{c}^{a}a_{\mu\nu}D_\mu^{ab}D_\nu^{bc}c^c+
\overline{c}^ia_{\mu\nu}\partial_\mu\left(\partial_\nu{c}^i+
gf^{abi}A_\nu^ac^b\right)+g\bar{c}^af^{abi}a_{\mu\nu}D_\mu^{bc}A_\nu^cc^i\nonumber\\
&+&g\bar{c}^aa_{\mu\nu}D_\mu^{ab}\left(f^{bcd}A_\nu^cc^d\right)-g^2f^{abi}f^{cdi}\bar{c}^ac^da_{\mu\nu}A_\mu^bA_\nu^c-
\frac{\alpha}{4}g^2f^{abi}f^{cdi}\bar{c}^a\bar{c}^bc^cc^d\nonumber\\
&-&\frac{\alpha}{4}g^2f^{abc}f^{adi}\bar{c}^b\bar{c}^cc^dc^i-\frac{\alpha}{8}g^2f^{abc}f^{ade}\bar{c}^b
\bar{c}^cc^dc^e+kgf^{abi}a_{\mu\nu}A_\mu^a\partial_{\nu}c^i\bar{c}^b\nonumber\\
&+&\left.kg^2f^{abi}f^{cdi}\bar{c}^ac^da_{\mu\nu}A_\mu^bA_\nu^c+
kg^2f^{abi}f^{bcj}a_{\mu\nu}A_\mu^iA_\nu^a\bar{c}^cc^j+
kgf^{abi}a_{\mu\nu}A_\mu^iD_\nu^{ac}c^c\bar{c}^b\right.\nonumber\\
&+&\left.kg^2f^{abi}f^{acd}a_{\mu\nu}A_\mu^iA_\nu^cc^d\bar{c}^b\right\}\;.\label{n66}
\end{eqnarray}
This expression coincides with the gauge fixing introduced in
\cite{Capri:2005zj}, which interpolates between the Coulomb,
Landau and MAG as well.  In order to achieve the real MAG
condition, {\it i.e.} $D^{ab}_{\mu}A^b_{\mu}=0$, the limit
$\alpha\rightarrow0$ has to be taken, namely
\begin{eqnarray}
S_{class}\bigg|_{\alpha\rightarrow0}&=&\int{d^4}x\left\{b^{a}\left[a_{\mu\nu}\partial_\mu^aA_\nu^a+
(k-1)gf^{abi}a_{\mu\nu}A_\mu^iA_\nu^b\right]+b^ia_{\mu\nu}\partial_\mu{A}_\nu^i+
\bar{c}^{a}a_{\mu\nu}D_\mu^{ab}D_\nu^{bc}c^c+\right.\nonumber\\
&+&\overline{c}^ia_{\mu\nu}\partial_\mu\left(\partial_\nu{c}^i+
gf^{abi}A_\nu^ac^b\right)+g\bar{c}^af^{abi}a_{\mu\nu}D_\mu^{bc}A_\nu^cc^i-
g\bar{c}^aa_{\mu\nu}D_\mu^{ab}\left(f^{bcd}A_\nu^cc^d\right)\nonumber\\
&-&g^2f^{abi}f^{cdi}\bar{c}^ac^da_{\mu\nu}A_\mu^bA_\nu^c+
kgf^{abi}a_{\mu\nu}A_\mu^a\partial_{\nu}c^i\bar{c}^b+kg^2f^{abi}f^{cdi}\bar{c}^ac^da_{\mu\nu}A_\mu^bA_\nu^c\nonumber\\
&+&\left.kg^2f^{abi}f^{bcj}a_{\mu\nu}A_\mu^iA_\nu^a\bar{c}^cc^j+
kgf^{abi}a_{\mu\nu}A_\mu^iD_\nu^{ac}c^c\bar{c}^b+kg^2f^{abi}f^{acd}a_{\mu\nu}A_\mu^iA_\nu^cc^d\bar{c}^b\right\}\;.
\nonumber\\
 & &\label{n6}
\end{eqnarray}
The gauge fixing conditions which stem from expression (\ref{n6})
are now easily seen to be derived by requiring that the following
field functional
\begin{equation}
\mathcal{F}=\int{d^4x}\frac{1}{2}a_{\mu\nu}\left(A_\mu^aA_\nu^a+kA_\mu^iA_\nu^i\right)\;.\label{func3}
\end{equation}
is stationary under the action of infinitesimal gauge
transformations. This requirememnt gives precisely the gauge
fixing conditions corresponding to (\ref{n6}), {\it i.e.}
\begin{eqnarray}
a_{\mu\nu}\left(D_\mu^{ab}A^b_\nu+kgf^{abi}A^b_\mu{A^i_\nu}\right)&=&0\;,\nonumber\\
a_{\mu\nu}\partial_\mu{A}_\nu^i&=&0\;.
\end{eqnarray}\\\\ One sees thus that a suitable minimizing functional can be associated to
the interpolating gauge (\ref{n6}), providing thus a useful way to
implement it on the lattice.

\end{document}